\documentclass[preprint, 5p, twocolumn]{elsarticle}

\usepackage{lineno,hyperref}
\modulolinenumbers[5]
\journal{*}

\usepackage{amsmath}
\usepackage{amsfonts}
\usepackage{algorithmic}
\usepackage[linesnumbered,ruled,vlined]{algorithm2e}
\usepackage{stfloats}
\usepackage{multirow}
\usepackage{multicol}
\usepackage{pgfplots}
\usepackage{tikz}
\usepackage{tablefootnote}
\usepackage{array}
\usepgfplotslibrary{statistics}
\usepackage{threeparttable}
\usepackage{layouts}
\usepackage{makecell}

\let \bs \boldsymbol

\def \xleft {\bs{x}_l}
\def \xright {\bs{x}_r}
\def \yleft {{y}_l}
\def \yright {{y}_r}

\def \figref {Figure}
\def\figpath {.}

\begin{document}

\begin{frontmatter}

\title{A Symmetric Regressor for MRI-Based Assessment of Striatal Dopamine Transporter Uptake in Parkinson's Disease With Enhanced Uncertainty Estimation}

\author[one]{Walid Abdullah Al}
\author[one]{Il Dong Yun\corref{mycorrespondingauthor}}
\cortext[mycorrespondingauthor]{Corresponding author}\ead{yun@hufs.ac.kr}

\address[one]{Division of Computer Engineering,\\ Hankuk University of Foreign Studies, Yongin, South Korea}

\author[two]{Yun Jung Bae}

\address[two]{Departments of Radiology, Seoul National University 
Bundang Hospital,\\ Seoul National University College of Medicine, Seongnam, Republic of Korea} 

%

\begin{abstract}

Dopamine transporter (DAT) imaging is commonly used for monitoring Parkinson's disease (PD), where the amount of striatal DAT uptake serves as the PD severity indicator. However, DAT imaging has a high cost and the risk of radiance exposure, besides its unavailability in general clinics. Recently, MRI of the nigral region has emerged as a safer and easier alternative. This work introduces a symmetric regressor for predicting the DAT uptake amount from the nigral MRI patch. Acknowledging the lateral symmetry between the right and left nigrae, the proposed regressor incorporates a paired input-output model that concurrently predicts the DAT uptake amounts for both the right and left striata. Moreover, it employs a symmetric loss that constrains the difference between right-to-left predictions, resembling the inherent correlation in DAT uptake amounts on the two lateral sides. Additionally, this work proposes a symmetric Monte-Carlo (MC) dropout method for providing a fruitful uncertainty estimate about the DAT uptake prediction, which utilizes the above symmetry. We evaluated the proposed approach on $734$ nigral patches, demonstrating significantly improved performance of the symmetric regressor compared with the standard regressors while giving better explainability and feature representation. Furthermore, the symmetric MC dropout gave precise uncertainty ranges for model predictions with a high probability of including the true DAT uptake amount within the range.

\end{abstract}


\end{frontmatter}


\section{Introduction}
\label{sec:introduction}
Parkinson's disease (PD) is one of the most common neurodegenerative syndromes \cite{bloem2021parkinson}, where a progressive loss of dopamine-containing neurons occurs in the midbrain \cite{damier1999substantia}. Dopamine transporter (DAT) imaging using single photon emission computed tomography (SPECT) can assess the degree of cell-loss based on the DAT uptake amount in the striatum of the midbrain \cite{wakabayashi2018semi}. Therefore, DAT imaging is widely used for diagnosing and monitoring the severity of PD \cite{suwijn2015diagnostic}. However, SPECT poses the risk of radiance exposure and is not allowed for patients with certain conditions (e.g., pregnancy) where administering radiotracers is not preferred \cite{akdemir2021dopamine}. Moreover, SPECT is only available in specialized clinics and has a high cost.

Recent studies suggested using MRI as a less invasive and readily available alternative imaging method for monitoring PD \cite{mahlknecht2017meta, schwarz2014swallow}. In MRI-based assessment, the loss of nigral hyperintensity is generally used for PD diagnosis as it corresponds well with the low DAT uptake. Nonetheless, a quantified severity assessment is not achievable by only manual observation of such intensity. Therefore, a computerized method for predicting the DAT uptake amount (as measured in SPECT) using the nigral MRI can be substantially useful. The Deep learning-based regressor showed promising results for DAT uptake prediction \cite{bae2023deep}. However, standard deep regressors generally minimizes the mean squared error (MSE) loss for the individual samples, where the samples are considered independent and identically distributed (i.i.d). On the contrary, the DAT uptake amounts in the right and left lateral sides have high correlation. Thus, the typical regressors fail to utilize such correlation.

\begin{figure*}[t]
\centering
\includegraphics[scale=1.0]{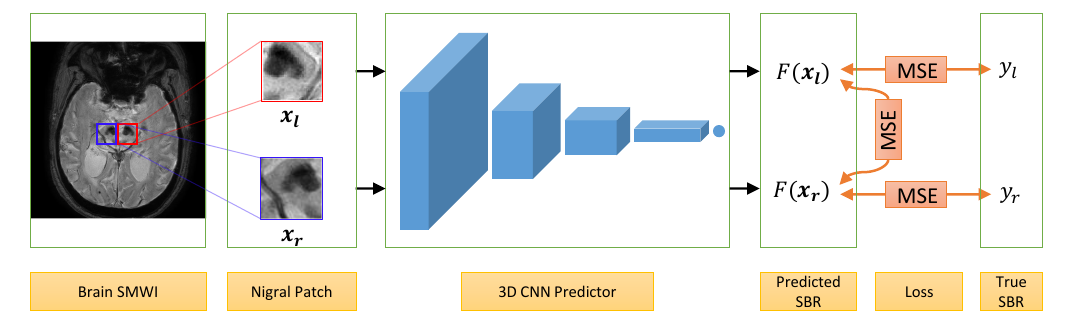}
\caption{{\bf The proposed symmetric regressor model.} Both the right and left SBRs are predicted simultaneously. Besides the regular MSE loss w.r.t. the true SBRs, the symmetric loss i.e., MSE between the right and left predictions is also minimized.}
\label{fig:method}
\end{figure*}

In this paper, we propose a symmetric deep regressor for predicting the DAT uptake amount from the MRI patch covering the nigral region. While the typical regressor follows a one-input-one-output structure with a loss function based on the mean squared error w.r.t. the known true output, our symmetric model adopts a paired input-output structure. Our regressor takes both the right and left nigral patches, simultaneously generating DAT uptake predictions for both the right and left striata. Additionally, we employ a symmetric loss that penalizes high differences between right and left predictions, addressing the symmetry between the two nigral regions and the strong correlation between DAT uptakes in the right and left striata. The proposed formulation facilitates explicit knowledge sharing between the two prediction tasks, thereby, allowing for enhanced feature learning. \figref~\ref{fig:method} illustrates the proposed symmetric regressor model. 

To improve the reliability of the proposed method during clinical use, we also propose a symmetric uncertainty estimation method about the generated predictions based on Monte-Carlo (MC) dropout \cite{gal2016dropout}. In our uncertainty estimation method, we again utilize the above symmetric relation between the right and left predictions to achieve improved uncertainty estimates. Therefore, our contribution is two-fold:
\begin{itemize}
\item We propose a symmetric regressor with a novel loss function for the striatal DAT uptake prediction, which utilizes the lateral symmetry and correlation of the right and left nigral patch inputs and the corresponding DAT uptake outputs.

\item We propose a symmetric MC dropout technique for improved uncertainty estimation, which enhances the reliability of the proposed method.
\end{itemize}
The code for our proposed method is available on Github at \url{https://github.com/awjibon/mri_dat/releases/tag/v1.0.0}

The remainder of the paper is organized as follows. In Section~\ref{sec:related_works}, we discuss the related works and contrastively point out the differences of the current work. Next, we describe our proposed methods in Section~\ref{sec:methods}. Afterwards, we present our experimental evaluation in Section~\ref{sec:results}. Finally, Section~\ref{sec:conclusion} presents our concluding remarks. 

\section{Related Works}
\label{sec:related_works}
\subsection{Standard DAT Uptake Assessment}
The degree of dopaminergic denervation is widely assessed using DAT imaging (i.e., $^{123}$I-FP-CIT SPECT) \cite{wakabayashi2018semi}. The disease progression is determined in terms of the specific binding ratio (SBR), which is obtained by comparing the counts of the striatal regions against the occipital cortex region \cite{rahman2019count}. Several works used the SBR for general PD diagnosis and progression tracing. For example, Kuo et al. \cite{kuo2019optimization} presented an SBR-based analysis for estimating PD progression with a one-year time window. Suwjin et al. \cite{suwijn2015diagnostic} evaluated a wide range of existing works in their review, which showed that SBR obtained from DAT imaging could successfully detect nigrostriatal dopaminergic cell loss with high accuracy.

Additionally,  a wide range of works also indicated the usefulness of SBR for early PD diagnosis. Filippi et al. \cite{filippi2005123i} utilized DAT imaging for detecting preclinical dopaminergic loss in PD. Booij et al. \cite{booij1997123i} showed an evident decline in striatal DAT uptake in early and advanced Parkinsonism. Prashanth et al. \cite{prashanth2014automatic} proposed an automatic classification model for classifying early PD based on the SBR.

\subsection{DAT Uptake Prediction using MRI} 
Recently, several studies addressed the need for an MRI-based functionality assessment because SPECT imaging has the risk of radiance exposure. Primarily, such studies investigated the use of nigral hyperintensity for PD diagnosis, concluding that the hyperintensity of nigrosome is usually lost in PD cases . Lehericy et al. \cite{lehericy2017role} suggested high-field MRI for PD imaging. Mahlknecht et al. \cite{mahlknecht2017meta} examined the hyperintensity of the nigral complex in MRI and used this as a marker for PD diagnosis. Schwarz et al. \cite{schwarz2014swallow} focused on the nigrosome's appearance in 3T MRI to determine PD. Gao et al. \cite{gao2015visualization} also emphasized the importance of nigrosome visualization in 3T MRI. They used the absence of nigrosome as a marker for PD identification. They utilized the susceptibility-weighted imaging (SWI) technique of MRI to enhance the visualization. Reiter et al. \cite{reiter2015dorsolateral} also used 3T SWI to determine the nigral hyperintensity effectively for PD classification. Cosottini et al. \cite{cosottini2015comparison} compared 3T and 7T SWI to investigate the substantia nigra for identifying PD. 

The use case of MRI is not limited to only PD diagnosis. Researchers have also used the nigral status in MRI to correlate with the SBR in SPECT.
Bae et al. \cite{bae2016loss} compared the nigral hyperintensity observed in 3T SWI with the SPECT SBR in Parkinsonism. In their later work \cite{bae2018loss}, they used a similar SWI-to-SBR comparison approach for idiopathic rapid eye movement disorder. Both of these works revealed good agreement between the SWI observation and the SBR score. Uchida et al. \cite{uchida2020magnetic} used the mean intensity of the manually annotated striatum region in order to correlate to the SBR score. Recently, susceptibility map-weighted imaging (SMWI) \cite{gho2014susceptibility} has been proposed for improving the image quality by providing with better contrast in neuroimaging. Nam et al. \cite{nam2017imaging} utilized SMWI to enhance the visibility of nigrosome for PD diagnosis, which allowed for better identification of the hyperintensity. Bae et al. \cite {bae2021imaging, bae2021determining} also compared SWI and SMWI images of the substantia nigra in Parkinsonism, and concluded that SMWI provides improved quality. 

The above MRI-based studies showed an overall good potential as an effective alternative. However, the primary objective of such studies was either to diagnose (not monitor) PD or to analyze a binary agreement between the nigral hyperintensity and the SBR. Such studies simply showed the relation of the presence (or absence) of hyperintensity to the normal (or abnormal) SBR findings. Nonetheless, the assessment of nigral hyperintensity is susceptible to the observer as it is manually evaluated. Moreover, such a binary assessment is also not helpful for PD monitoring. Manual quantification of the hyperintensity to correlate with the continuous SBR measures is also infeasible. Therefore, a computational method for quantifying the hyperintensity while maintaining a high correlation with SBR is necessary.

\subsection{Deep Learning Based Prediction}
Convolutional neural networks (CNNs) have been widely used for processing spatial inputs such as X-ray images, MRI volumes, etc. Though CNNs (e.g., VGG \cite{simonyan2014very}, ResNet \cite{he2016deep}) were originally proposed for natural image analysis in the general computer vision domain, they have also shown promising outcomes in analyzing medical images. Litjens et al. \cite{litjens2017survey} presented a comprehensive survey where they included diverse medical image analysis works based on such deep convolutional networks.

While such networks are mainly used for classification problems (e.g., determining tumor categories \cite{mohsen2018classification}, identifying healthy and diseased cases \cite{yadav2019deep}, etc.), CNN-based medical applications for predicting continuous values are relatively few. Examples of such continuous-valued outputs include disease severity, fetus head circumference, etc. Aboutalebi et al. presented \cite{aboutalebi2021covid} a CNN-based model for assessing severity in COVID-19 from X-ray images. Zhang et al. adopted a CNN model \cite{zhang2020direct} for automatically measuring the fetus's head circumference in ultrasound images.

Training a CNN model for such prediction is generally formulated as a supervised regression problem where the MSE loss between the model prediction and the desired output is minimized. In our previous work \cite{bae2023deep}, we also applied such a CNN regression model for predicting the SBR from SMWI patches of the nigra. However, we only focused on estimating the left striatal SBR based on the left nigral patch. Most regression CNNs for medical tasks are simply adopted by the plug-and-play method, where the data points (input images and the corresponding outputs) are considered independent and identically distributed (i.i.d). However, directly applying such models in the medical domain is not always optimal as it fails to utilize the underlying data correlation resulting from the anatomy.

In this work, we utilize both the right and left nigral patches to predict SBRs of both the right and left striata. Leveraging the high correlation between the left and right striatal SBRs and the symmetric structure of the two nigrosomes along with their surrounding regions, we propose an improved regression model with a novel loss function (\figref~\ref{fig:method}). Furthermore, we introduce an uncertainty estimation mechanism based on the symmetry for reliable use in clinical sites.

\section{Methodology}
\label{sec:methods}
\subsection{Data}
We collected the dataset for this study from Seoul National University Bundang Hospital, South Korea. Approval of all ethical and experimental procedures and protocols was granted by the Institutional Review Board (IRB) of Seoul National University Bundang Hospital, South Korea (IRB number: B-1610-368-303). Written informed consent was obtained from 
all study participants.

We aim to predict the striatal SBR values using the 3D nigral patches obtained from SMWI. Each case involves two target SBR values for the right and left striata. Therefore, the right nigral patch is used for predicting the right striatal SBR and the left nigral patch is used for the SBR of the right striatum. From the original SMWI volume of the brain, a $50 \times 50 \times 20$ voxels region centered at the nigrosome is handcropped by the physicians to prepare the nigral patch. The above patch size adequately accommodates the substantia nigra, including the nigrosome and red nucleus. Note that the red nucleus is also affected by PD \cite{rodriguez2008neuronal}. We denote the input nigral patches by $\xright$ and $\xleft$ and the target SBRs by $\yright$ and $\yleft$.

\subsection{Deep Regressor}
Deep regressors with spatial input usually utilize CNN, where a series of convolutional blocks is used to extract spatial features. These features are then passed to a fully connected layer to obtain the final prediction. We also use a CNN regressor to model our SBR regression function $F$. Because our input is a 3D patch, we incorporate a 3D CNN where 3D convolutional blocks are used using a VGG-style architectural organization. However, we reduce the number of layers to fit our small input dimension. Table~\ref{tab:vgg} presents our proposed VGG.

\begin{table}[t]
\caption{Proposed VGG Architecture for SBR Regression.}
\label{tab:vgg}
\renewcommand{\arraystretch}{1.2}
\begin{center}
\begin{tabular}{l c c}
\hline
{\bf Layer} & {\bf No. of kernels,} & {\bf Output size}\\
&  {\bf kernel size \textbackslash stride}\\
\hline
Input & $-$ & $50 \times 50 \times 20 \times 1$\\
Conv1-1 & $64, 3 \times 3 \times 3$ \textbackslash $1$ & $50 \times 50 \times 20 \times 64$\\
Conv1-2 & $64, 3 \times 3 \times 3$ \textbackslash $1$ & $50 \times 50 \times 20 \times 64$\\ 
Pool1 & $-, 2 \times 2 \times 2$ \textbackslash $2$ & $25 \times 25 \times 10 \times 64$\\ 

Conv2-1 & $128, 3 \times 3 \times 3$ \textbackslash $1$ & $25 \times 25 \times 10 \times 128$\\
Conv2-2 & $128, 3 \times 3 \times 3$ \textbackslash $1$ & $25 \times 25 \times 10 \times 128$\\ 
Pool2 & $-, 2 \times 2 \times 2$ \textbackslash $2$ & $12 \times 12 \times 5 \times 128$\\ 

Conv3-1 & $256, 3 \times 3 \times 3$ \textbackslash $1$ & $12 \times 12 \times 5 \times 256$\\
Conv3-2 & $256, 3 \times 3 \times 3$ \textbackslash $1$ & $12 \times 12 \times 5 \times 256$\\ 
Pool3 & $-, 2 \times 2 \times 2$ \textbackslash $2$ & $6 \times 6 \times 2 \times 256$\\ 

Conv4-1 & $512, 3 \times 3 \times 3$ \textbackslash $1$ & $6 \times 6 \times 2 \times 512$\\
Conv4-2 & $512, 3 \times 3 \times 3$ \textbackslash $1$ & $6 \times 6 \times 2 \times 512$\\ 
Pool4 & $-, 2 \times 2 \times 2$ \textbackslash $2$ & $3 \times 3 \times 1 \times 512$\\

Flatten & $-$ & $4608$\\
FC & $-$ & $1$\\
\hline
\end{tabular}
\end{center}
Conv: convolutional layer, Pool: max-pooling layer, FC: fully connected layer.\\
All the convolutional layers used zero-padding.
\end{table}

During training, the model is optimized to minimize the MSE between the model predictions and true SBRs about the training examples. Hence, the loss function is as follows:
\begin{equation}
L(F) = \mathbb{E}_{(\bs{x}, y)} \big[ \big(y - F(\bs{x})\big)^2\big]
\end{equation}
where $\bs{x}$ is the model input and $y$ is the target true value.

Our problem has two inputs and two corresponding outputs. This can be modeled in two ways: (i) use separate models for the right and left striatal SBR predictions, and (ii) share a single model for both. We adopted the former option in our previous work on left striatal SBR prediction \cite{bae2023deep}. Nonetheless, this option ignores any correlation between the right and left data. In this work, we suggest incorporating a shared model. Subsection~\ref{sec:symreg} elaborates on the advantage of this approach.
The regression loss function for using  such a shared model can be expressed as follows:
\begin{equation}
\begin{split}
L_\text{reg}(F) =  { } &\mathbb{E}_{(\xright, \yright)} \big[ \big(\yright - F(\xright)\big)^2\big]+\\
& \mathbb{E}_{(\xleft, \yleft)} \big[ \big(\yleft - F(\xleft)\big)^2\big]
\end{split}
\label{eq:loss}
\end{equation}

\subsection{Symmetric Regressor}
\label{sec:symreg}
\subsubsection{Identical Input Type}
In our symmetric regressor, we utilize the symmetric property of the substantia nigra. Like other paired organs in the human body, the right and left nigrae appear to be mirrored versions of one another. The surrounding structures in the mid-brain also follow this symmetry. Therefore, we formulate right and left nigral patches as identical input types by modeling the right nigra as the laterally flipped left nigra. By doing so, we can use a single model for both SBR predictions, where right and left patches can simultaneously be used for training without any distinction. 

The above formulation potentially gives a couple of advantages. Firstly, it reduces the number of model parameters because only a single prediction model can be used. Secondly, training data is increased because both the right and left striatal data can be used simultaneously. Thirdly, it enables improved feature learning by implicit knowledge sharing between the right and left SBR prediction tasks. 

\subsubsection{Correlated Output}
\begin{figure}
\centering
\includegraphics[scale=1.0]{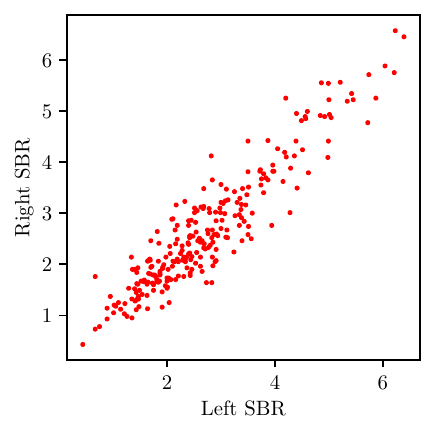}
\caption{{\bf R-R plot of our right and left SBRs showing a correlation of $0.93$}.}
\label{fig:lr_sbr}
\end{figure}

Furthermore, we also argue that the SBR scores of the right and left striata are also connected. Usually, dopaminergic cell loss occurs similarly on both sides. Consequently, a low SBR score on the left striatum is likely to indicate a low SBR score on the right, and vice versa. Empirically, the right and left SBRs gave a high correlation coefficient of $0.93$ based on our data. \figref~\ref{fig:lr_sbr} presents the correlation plot of the right and left SBR values. We propose to enforce an explicit constraint regarding this relation in our symmetric regressor.

\subsubsection{Proposed Model}
For effectively utilizing the above correlations, we propose a symmetric model $F_\text{sym}$ that simultaneously takes a pair of inputs (i.e., the right and left nigral patches) and produces a pair of outputs (i.e., the right and left SBRs). Inside this symmetric model, a single CNN regressor $F$  processes both inputs in the pair. Formally, 
\begin{equation}
F_\text{sym} (\xright, \xleft) = \big(F(\xright), F(\xleft)\big) 
\end{equation}
The MSE loss for training the above model could be defined as follows:
\begin{equation}
\begin{split}
L(F_\text{sym}) &= \mathbb{E}_{(\xright, \xleft, \yright, \yleft)} \big[ | (\yright, \yleft) - (F(\xright), F(\xleft))|^2\big]\\
&= \mathbb{E}_{(\xright, \xleft, \yright, \yleft)} \big[ \big(\yright-F(\xright)\big)^2 +  \big(\yleft-F(\xleft)\big)^2\big]\\
&= L_\text{reg}(F)
\end{split}
\end{equation}
which is essentially the same loss function as \eqref{eq:loss}. This indicates the same effect as training with all the right and left data with a single input model. However, the proposed paired input approach is required for our latter addition about the output constraint.

We utilize the output correlation by imposing a constraint that the predicted right and left SBRs should be close to each other. We do so by adding a loss term incorporating the difference between the right and left predictions. Specifically, the proposed loss term is as follows:
\begin{equation}
L_{\text{sym}}(F) = \mathbb{E}_{(\xright, \xleft)} \big[ \text{clip} \big(\big(F(\xright) - F(\xleft)\big)^2, \alpha^2\big)\big]
\end{equation}
where
\begin{equation}
\text{clip}(a, \alpha^2) = 
\begin{cases}
a &\text{if} \;  a > \alpha^2\\
0 &\text{otherwise}
\end{cases}
\end{equation}
This clipping is required because the right and left SBRs are still not the same despite being closely related. They still have a small difference that we should consider in this additional loss term. $\alpha$ controls the margin, below which the right-to-left difference is ignored. We choose the optimal value for $\alpha$ by experiment.

Therefore, our final loss function becomes a weighted sum of the two losses, which can simply be stated as follows:
\begin{equation}
L_\text{final}(F) = L_\text{reg}(F) + \beta L_\text{sym}(F) 
\end{equation} 

\subsection{Symmetric Uncertainty in Prediction}
The MRI-based SBR prediction is a relatively new problem and has not yet been established as a regular medical practice. Therefore, estimating uncertainty about the predicted SBRs can be crucial for the practical application of the proposed model. Usually, uncertainty in regression is expressed through prediction intervals, which provide a range around the predicted SBR within which the true SBR is likely to fall. 

\subsubsection{Monte-Carlo Dropout:}
MC dropout \cite{gal2016dropout} is a widely used uncertainty estimation technique for deep regressors. The dropout layer is useful for improving generalization and preventing overfitting during training neural networks. The dropout operation indicates randomly dropping some neurons  (i.e., randomly making some neurons zero). Usually, the dropout layer is only used in training and deactivated during testing. However, the MC dropout technique suggests keeping the dropout layer even in the testing phase.

Due to the active dropout, the model gives a different prediction after each traversal for a single test input. Thus after $n$ runs, a population of $n$ different prediction results can be obtained for each test case. The prediction interval (i.e., uncertainty in prediction) is then calculated as the standard deviation (SD) of the $n$ prediction results. The wider the SD, the higher the uncertainty.

\subsubsection{Symmetric MC dropout:}
We also follow the MC dropout-based interval estimation for expressing the uncertainty in SBR prediction. Unlike the usual interval estimation method, we utilize the predictions of both the right and left SBR predictions to determine the interval.

Because our symmetric regressor takes both the right and left nigral patches of a patient and produces the right and left SBRs simultaneously, the MC dropout technique would give two distributions. Let us suppose that we test our symmetric regressor for some input-pair $(\xright, \xleft)$. After $n$ test runs, the observed outputs are $(\hat{\yright^1}, \hat{\yleft^1}), ..., (\hat{\yright^n}, \hat{\yleft^n})$. Hence, the SDs of the right and left observations are as follows:
\begin{equation}
\begin{split}
\sigma_r(\xright) &= \sqrt{\frac{1}{n} \sum_{t=1}^n \big( \hat{\yright^t} - \bar{\yright}\big)^2}\\
\sigma_l(\xleft) &= \sqrt{\frac{1}{n} \sum_{t=1}^n \big( \hat{\yleft^t} - \bar{\yleft}\big)^2}\\
\end{split}
\end{equation}
where $\bar{\yright}$ and $\bar{\yleft}$ are the observation-means for the right and left outputs, respectively. The existing MC dropout suggests using these two independent SDs for the corresponding predictions i.e., $\sigma_r$ indicating the interval/uncertainty about the right SBR and $\sigma_l$ indicating the interval about the left. 

In our uncertainty formulation, we incorporate both predictions $(\hat{\yright^t}, \hat{\yleft^t})$ for determining the interval about either of the right or left SBR predictions, owing to their correlation as argued before. Similar to the symmetric loss, we consider the pairwise differences of the right and left MC predictions and take their mean as an indication of the prediction interval. Formally, the suggested interval hint is as follows:
\begin{equation}
\sigma_\text{sym}(\xright, \xleft) = \frac{1}{n}\sum_{t=1}^n |\hat{\yright^t}-\hat{\yleft^t}|
\end{equation}
Thus, we hypothesize that a large difference in the right and left outputs indicates a probable large uncertainty in the model prediction. Therefore, we inject this uncertainty hint into the original MC intervals to improve model reliability. We define our final prediction intervals as follows:
\begin{equation}
\begin{split}
\sigma_r^{\text{sym}}(\xright, \xleft) = \sigma_r (\xright) + \gamma_l \sigma_\text{sym} (\xright, \xleft)\\
\sigma_l^{\text{sym}}(\xright, \xleft) = \sigma_l (\xleft) + \gamma_r \sigma_\text{sym} (\xright, \xleft)
\end{split}
\end{equation}
by combining the symmetric hint into the MC intervals. Here, $0 \le \gamma_r, \gamma_l \le 1$ are chosen empirically from the validation set. Algorithm~\ref{alg:sym_mc} presents our prediction interval computation method.

\begin{algorithm}[t]
\caption{Symmetric MC Prediction Interval}
\label{alg:sym_mc}
\begin{algorithmic}
\STATE \textbf{Input:} Predictor $F_\text{sym}$ with dropout layer activated; right and left nigral patches $\xright, \xleft$; symmetric interval weights $\gamma_r, \gamma_l$.
\STATE \textbf{Output:} right \& left SBR prediction intervals $\sigma_r^\text{sym}, \sigma_l^\text{sym}$.
\STATE \textbf{Initialize} MC predictions $Y=\emptyset$.
\FOR {$t=1$ to $n$}
	\STATE Get predictions  $(\hat{y_r^t}, \hat{y_l^t}) = F_\text{sym}(\xright, \xleft)$.
	\STATE Append $(\hat{y_r^t}, \hat{y_l^t})$ to $Y$.
\ENDFOR
\STATE \textbf{with} gathered predictions $(\hat{y_r^t}, \hat{y_l^t}) \in Y$,
\STATE Get mean pairwise difference $\sigma_\text{sym}$=$\text{mean}\big((|\hat{y_r^t}-\hat{y_l^t}|)_{t=1}^n\big)$
\STATE Get MC intervals $(\sigma_r, \sigma_l) = \big(\text{SD}\big((\hat{y_r^t})_{t=1}^n\big), \text{SD}\big((\hat{y_l^t})_{t=1}^n\big) \big)$
\STATE Get final intervals $(\sigma_r^\text{sym}, \sigma_l^\text{sym})=(\sigma_r + \gamma_r \sigma_\text{sym}, \sigma_l + \gamma_l \sigma_\text{sym})$
\end{algorithmic}
\end{algorithm}

\section{Experimental Results}
\label{sec:results}
\subsection{Experimental Setting}
For experimental evaluation, we acquired $734$ nigral patches (right and left combined) of $367$ patients ($55.33\%$ female; $69.0 \pm 9.2$ years old) from Seoul National University Bundang Hospital, South Korea. Among the $367$ patiens, the number of PD patients were $287$. Note that, this dataset was previously used in Bae et al.'s work \cite{bae2023deep}. However, only left nigral patches were used in their work. The nigral SMWI patch and SPECT SBR information of our dataset is presented in Table~\ref{tab:data}. The SMWI patches are obtained from a 3T MRI scanner. The detailed protocols of MRI scanning, SMWI image reconstruction, and DAT SPECT imaging are described in \cite{bae2023deep}. Two neuroradiologists (with 12 and 22 years of experience) annotated the nigrosome centroids in consensus for cropping the patches. The true SBR scores were automatically obtained from the SPECT images by the DATquant program by GE Healthcare.

\begin{table}[!t]
\caption{Data Description}
\label{tab:data}
\renewcommand{\arraystretch}{1.2}
\centering
\begin{tabular}{l l}
\hline
{\bf Attribute} & {\bf Description}\\
\hline
\hline
Total cases & $734$ ($367$ right and $367$ left \\
& nigral patches)\\
\cline{1-2}
Train:validation:test & $512$:$112$:$110$ ($70\%$:$15\%$:$15\%$)\\
\cline{1-2}
Data acquisition period & February $2017$-December $2018$\\
\cline{1-2}
MRI (SMWI) device & Ingenia and Ingenia CX, \\
& Philips, Netherlands. Detail \\
& protocol in\cite{bae2023deep}\\
\cline{1-2}
Nigral patch size & $50\times 50\times 20$ voxels\\
\cline{1-2}
Voxel size & $0.5 \times 0.5 \times 1$ mm$^3$\\
\cline{1-2}
SPECT device & DATrace-123™, Samyoung \\ 
& Unitech, Korea with Trionix \\
& XLT, Trionix Research Lab,\\
&  USA. Detail protocol in \cite{bae2023deep}\\
\cline{1-2}
SBR analysis program & DATquant, Xeleris 3.1, GE \\
& Healthcare, USA\\
\cline{1-2}
SBR range & Right: $[0.44, 6.84]$;\\
&  left: $[0.43, 6.80]$\\
\cline{1-2}
MRI/SPECT time diff. & $42 \pm 60$ days\\
& (median: $18$ hours)\\
\hline

\end{tabular}
\end{table}

\begin{table*}[!t]
\caption{SBR Prediction Performance.}
\label{tab:error}
\renewcommand{\arraystretch}{1.2}
\begin{center}
\begin{tabular}{l |c c c| c c c|c c c}
\hline
 & \multicolumn{3}{c|}{\bf RMSE} & \multicolumn{3}{c|}{\bf MAE}  & \multicolumn{3}{c}{\bf $R$}\\
{\bf Methods} & Right & Left & Avg. & Right & Left & Avg. & Right & Left & Avg.\\
\hline
ResNet-$18$ \cite{he2016deep}& 1.3955 & 1.6411 & 1.5183 & 0.8836 & 0.9459 & 0.9147 & 0.4458 & 0.3255 & 0.3856 \\
VGG-$16$ \cite{simonyan2014very}& 1.5515 & 1.6699 & 1.6107 & 0.9829 & 1.0236 & 1.0032 & 0.3978 & 0.3112 & 0.3545 \\
VGG-small \cite{bae2023deep} & 0.9410 & 1.0553 & 0.9982 & 0.7687 & 0.7663 & 0.7675 & 0.6281 & 0.7067 & 0.6674 \\
Symmetric VGG* & 0.8477 & {\bf 0.6929} & 0.7703 & 0.7238 & 0.6738 & 0.6988 & 0.6745 & 0.7405 & 0.7075\\
Symmetric VGG+$L_\text{sym}$*  & {\bf 0.8180} & 0.7081 & {\bf 0.7630} & {\bf 0.6774} & {\bf 0.6709} & {\bf 0.6741} & {\bf 0.7157} & {\bf 0.7426} & {\bf 0.7291}\\
\hline
\multicolumn{1}{l}{* Proposed}\\
\end{tabular}\\
\end{center}
\end{table*}

After preparing the paired dataset, we performed a $70\%:15\%:15\%$ random split to obtain the training, validation, and testing set, respectively. The validation set was used to decide model hyperparameters while the test set was only used in the final evaluation phase. In this section, we first show the model dependency on different hyperparameters. Next, we present our evaluation results on the SBR prediction performance. Finally, we evaluate the uncertainty estimated by the proposed symmetric MC method.

\subsection{Hyperparameter Dependency}
\begin{figure}
\centering
\includegraphics[scale=1.0]{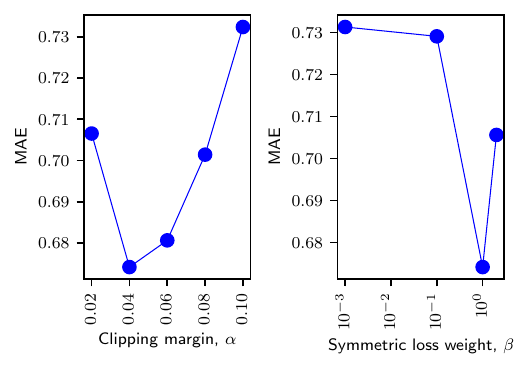}
\caption{{\bf Dependency on the clipping margin (left) and weight (right) of the symmetric loss.} The clipping margin in the left plot is normalized relative to the maximum SBR ($6.84$) of our dataset. The weight axis in the right plot is in log scale. }
\label{fig:hyperparameter}
\end{figure}

The key hyperparameters in the proposed model are the clipping margin $\alpha$ and weight $\beta$ of the symmetric loss. We present the mean absolute error (MAE) of the SBR prediction for different values of $\alpha$ and $\beta$ in \figref~\ref{fig:hyperparameter}. Note that the optimal hyperparameters were chosen based on the validation set performance. However, the dependency plots here are presented for the test set.

The clipping margin contributes to ignoring small right-to-left differences while computing the symmetric loss. From the resultant plot in \figref~\ref{fig:hyperparameter}(left), it is apparent that a small margin gave a poor performance because it strictly penalized the objective even for the small symmetric differences. On the other hand, a large margin also gave a non-optimal performance because it ignored most of the differences including the cases where the right-to-left prediction differences were big. With the optimal clipping margin, the loss only ignored (normalized) prediction differences less than $0.04$ while penalizing the objective for larger differences, thus improving the prediction performance.

The symmetric loss weight controls the influence of the symmetric loss over the original regression loss. We present the prediction performance for different weight values in \figref~\ref{fig:hyperparameter}(right). Note that the weight axis in the plot is presented in the logarithmic scale. In this experiment, we used the optimal clipping margin i.e., $\alpha=0.04$ (decided previously). A high weight puts too much focus on the symmetric loss resulting in a higher MAE. Contrarily, a low weight fails to effectively utilize the symmetric property of our input and output pairs.

\subsection{SBR Prediction Performance}
Following general continuous valued prediction problems, we evaluate the proposed regressor based on the MAE and root mean squared error (RMSE) between the model predictions and the true SBR values. Furthermore, we also compare Pearson's correlation coefficient $R$, which is a widely used performance measure in the medical field. We present the SBR prediction performance of different regressor models including ours in Table~\ref{tab:error}. We show the improvements made by our propositions in three steps. First, we compare the proposed VGG architecture ($VGG$-$small$) with the standard VGG-16 \cite{simonyan2014very} and ResNet-18  \cite{he2016deep} architectures.
Second, we present the improvements of the paired input-output model ($symmetric$ $VGG$). Finally, we compare the performance of the paired model trained with the symmetric loss ($symmetric$ $VGG$ + $L_{\text{sym}}$).

As Table~\ref{tab:error} demonstrates, both the VGG-16 and ResNet-16 models gave a poor prediction performance, resulting an MAE above $0.915$. The $VGG$-$small$ network gave a better prediction error and correlation than the widely used VGG-16 and ResNet-18 networks, improving the error to $0.768$. The paired input-output model gave a further improvement by combining the right and left prediction tasks. The final model trained with the symmetric loss generally outperformed all the other models by dragging the prediction error down to $0.6741$.

\subsection{Quality of Uncertainty}
\label{sec:res_uncertainty}
Evaluating the uncertainty (i.e., the prediction interval) also plays a significant role in the medical domain. A reliable medical application must have good precision while providing a prediction interval that would include the true SBR with a high probability. One of the most common evaluation measures of the regression uncertainty is the coverage probability (CP) \cite{psaros2023uncertainty} i.e., the probability that the true value lies in the specified interval. However, only using this metric is problematic because one can easily ensure a $100\%$ CP by using a substantially large interval despite that interval being useless. Therefore, we consider a second metric called $sharpness$ \cite{psaros2023uncertainty}, which is defined as $1.0  -$normalized width of the interval.

The CP and sharpness have a reciprocal relation much like the recall and precision measures in the classification task. At $0\%$ CP, we can ensure a $100\%$ sharpness. As the CP increases, the sharpness decreases. Finally, sharpness reaches at $0\%$ when CP $=100\%$. 
To effectively evaluate the regression uncertainty, we propose to plot the sharpness-CP curve (similar to the precision-recall curve) and calculate the area under the curve (AUC).

\begin{figure}
\centering
\includegraphics[scale=1.0]{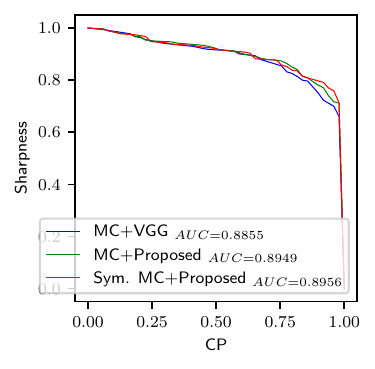}
\caption{{\bf Sharpness-CP curves for the prediction intervals of different methods.}}
\label{fig:auc}
\end{figure}

For plotting the curve, we obtain the (sharpness, CP) points by varying the interval from $0.\sigma$ to $m\sigma$, given the original interval $\sigma$ from MC dropout. Here, $m>0$ is defined such that the CP reaches $100\%$. \figref~\ref{fig:auc} presents the resultant sharpness-CP curves. This figure presents three curves given by the MC intervals of the standard VGG regressor, the MC intervals of the proposed regressor, and our symmetric MC intervals of the proposed regressor.

Both intervals of the proposed symmetric regressor had a greater AUC than the standard regressor. The interval we proposed using symmetric MC dropout gave a slightly better AUC than the usual MC dropout. For the final regressor, the curves of the symmetric MC and MC intervals seems almost similar except only in a few places. However, the proposed MC is still advantageous because high CP positions are usually required for practical application in medical sites. At high CP (e.g., above $80$\% CP), the symmetric MC gave intervals with higher sharpness and precision. We discuss more on this in Section~\ref{sec:uncertainty}.

\section{Discussion}
\subsection{Validity of the Proposed Model}
\subsubsection{CNN Model Selection}
As mentioned before, we did not use the standard CNN architectures for our CNN encoder in the symmetric regressor. Rather, we utilized a smaller version of the VGG-style architecture.
From Table~\ref{tab:error}, we observed that the proposed VGG network ($VGG$-$small$) gave a lower error than the widely used ResNet-18\cite{he2016deep} and VGG-16 \cite{simonyan2014very} networks. While ResNet-18 and VGG-16 networks were originally proposed for the ImageNet dataset consisting of 2D natural images, the 3D nigral patch has a much smaller spatial dimension. The proposed network gave improved performance because it had a smaller number of layers suitable for the small input.

\subsubsection{Advantage of the Symmetric Model}
Compared with the standard regressor with individual right and left predictors, our paired model performed significantly better. In the paired model, right and left SBRs were predicted simultaneously due to the lateral symmetry and correlation between the right and left nigral patches. As a result, the model was provided with more training data on top of implicit knowledge sharing between the highly correlated right and left prediction tasks. Thereby, it gave a better generalization utilizing the symmetric property. Moreover, the symmetric loss helped train the paired model even more effectively by enforcing an additional constraint based on the output correlation in the pairs. As a result, our final model ($Symmetric$ $VGG$ + $L_\text{sym}$) gave a further improvement overall.

The proposed symmetric regressor resulted in an average correlation of  $0.7291 $ ($p$ value $<0.000001$), which is a considerable increase from the standard regressor's correlation (i.e., $0.6674$, $p$ value $<0.000001$). Note that, this resultant correlation is within the acceptable range in the medical field. The resultant MAE is also improved from $0.7675$ to $0.6741$ by the proposed method. We compared the results of $VGG$-$small$, $Symmetric$ $VGG$, and $Symmetric$ $VGG$ + $L_\text{sym}$ using Student's $t$-test, which revealed that the improvements by $Symmetric$ $VGG$ (w.r.t $VGG$-$small$) and $Symmetric$ $VGG+ L_\text{sym}$ (w.r.t $Symmetric$ $VGG$) were statistically significant ($p$ value $<0.01$ and $p$ value $<0.03$, respectively).

\begin{table*}[!t]
\caption{Comparison with Other MRI-based DAT Uptake Analysis Works.}
\label{tab:comparison}
\renewcommand{\arraystretch}{1.2}
\centering
\begin{tabular}{l | c c | c c }
\hline
 & \multicolumn{2}{c|}{\bf Prediction Correlation} &\multicolumn{2}{c}{\bf Binary Agreement (acc., sen., spe.)}\\
{\bf Method; goal; data size } & {Right} & {Left} & {Right} & {Left}\\
\hline
 Mean intensity \cite{uchida2020magnetic}; classification; $61$ & \multicolumn{2}{c|}{$0.470$}& - & -\\
Observation \cite{bae2021determining}; classification; $138$ & - & - & ($0.652, 0.587, \bs{0.897}$) & ($0.667, 0.596, \bs{0.932}$)\\
VGG regressor \cite{bae2023deep}; prediction; $367$ & $0.628$ & $0.707$ & ($0.764, 0.721, 0.750$) & ($0.855, 0.786, 0.756$)\\
This work; prediction; $734$ & $\bs{0.716}$ & $\bs{0.743}$ & ($\bs{0.818}, \bs{0.860}, 0.833$) & ($\bs{0.873}, \bs{0.854}, 0.786$)\\
\hline
\end{tabular}
\end{table*}

\subsection{Comparison with Previous Works}
Most of the previous MRI-based SBR prediction works attempted to investigate a binary agreement between the normal/abnormal MRI and SPECT findings. For fruitful comparison, we also obtained such binary decisions from our continuous valued data. Based on the previous studies \cite{bae2021determining, bae2023deep}, we used a threshold of $3.401$ (right), $3.345$ (left) to determine the normal and abnormal SBRs. Thus, we compare our work with the previous MRI-based studies in Table~\ref{tab:comparison}. In this table, we present the correlation for the continuous valued prediction. Moreover, we depict the accuracy, sensitivity, and specificity for the binary decision (i.e., classification of normal and abnormal SBRs).

Among the SBR prediction studies, Uchida et al. \cite{uchida2020magnetic} with their mean striatal intensity approach could achieve a correlation of $0.470$, while the standard deep regressor of Bae et al. \cite{bae2023deep} gave a correlation of $0.667$. Following the standard regression approaches, the regressor in \cite{bae2023deep} was trained with the usual regression loss, attempting to minimize the individual sample-wise difference between the predicted and actual SBRs. Our proposed regressor utilizing the pairwise symmetric loss between the right and left SBRs could outperform the standard regressor while utilizing the same dataset as \cite{bae2023deep}.

Among the SBR classification studies, the manual identification approach of Bae et al. \cite{bae2021determining} gave an average specificity of $0.915$. However, the accuracy ($0.659$) and sensitivity ($0.591$) was substantially low. Note that a good sensitivity is important in the medical field as it signifies how many abnormal findings are mistakenly categorized as being normal. The deep regressor of Bae et al. \cite{bae2023deep} gave a well-balanced specificity and sensitivity with an accuracy of $0.809$. On the other hand, our symmetric regressor gave the highest accuracy ($0.845$). In our work, the average sensitivity and specificity was $0.857$ and $0.810$, thus resulting in a significantly better SBR classification performance ($p$ value $< 0.0004$, as revealed with McNemar's test).

\subsection{Explainability Analysis}
\begin{figure}
\centering
\includegraphics[scale=1.0]{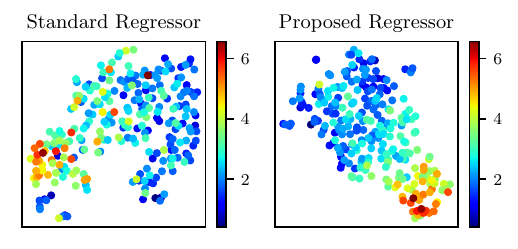}
\caption{{\bf t-SNE visualization of the extracted features by the standard and proposed regressors.} Each point is color-graded based on its corresponding true SBR value.} 
\label{fig:tsne}
\end{figure}

We also visually investigated the extracted features by the proposed model to evaluate the discriminative nature of the learned representation of our data. We did so by mapping the extracted features into 2D space using t-SNE \cite{van2008visualizing}. We plot the resultant 2D features in \figref~\ref{fig:tsne}. Each point in this plot is color-graded based on its true SBR value. We also plot the features extracted by the standard regressor for comparison. It is clear from the plots that the proposed model could learn comparatively more powerful feature representation that is useful for discriminating among different SBR ranges. 

Additionally, we analyzed the activation maps of the regressors to examine the region of focus responsible for SBR prediction. Among the recent methods, HiResCAM \cite{draelos2020use} is one of the most prevalent methods for obtaining activation maps for deep networks. Compared to the other methods, HiResCAM can generate higher-resolution maps. Nonetheless, HiResCAM, like all the other similar methods, is well-suited for classification tasks, where gradients with respect to a specific class score are used to obtain the responsible region for that specific class. To effectively incorporate HiResCAM in our regression model, we consider two classes of SBRs- high and low. We obtain the score/probability of the high SBR class as the normalized SBR prediction $\tilde{F}(\bs{x})=F(\bs{x})/SBR_{\text{max}}$. Moreover, the probability of the low SBR class is obtained by $\tilde{F}'(\bs{x}) = 1-\tilde{F}(\bs{x})$. Thus, the gradients of high SBR probability $\tilde{F}(\bs{x})$ point to the direction of increased SBR, and the gradients of low SBR score $\tilde{F}'(\bs{x})$ correspond to the decreasing direction of SBR. We distinguish between high and low classes using a normalized SBR threshold of $0.5$.

\begin{figure*}
\centering
\includegraphics[scale=1.0]{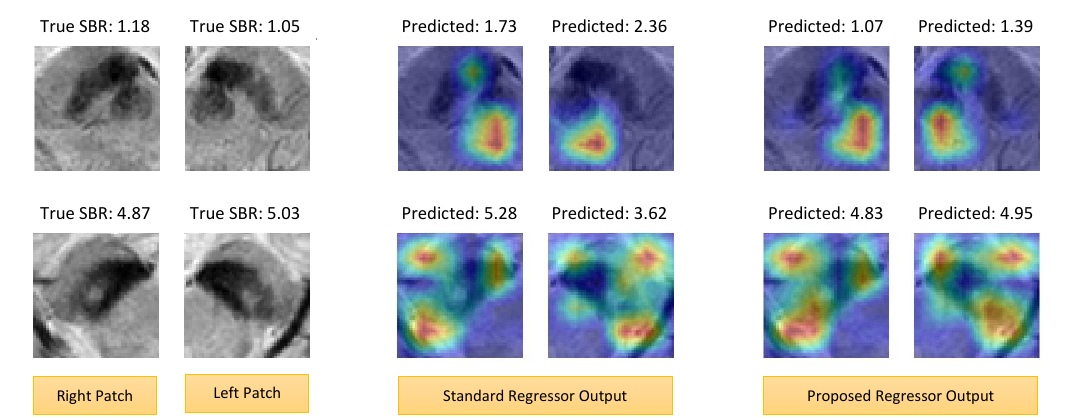}
\caption{{\bf Grad-CAM activation maps of the standard and proposed regressors.} (top) Low SBR case. (bottom) High SBR case. Overall, the heatmaps of the proposed method gave stronger attention and coverage of the target regions (e.g., nigrosome, substantia nigra, red nucleus, etc.).}
\label{fig:cam}
\end{figure*}

\figref~\ref{fig:cam} presents one example for each of the high and low SBR cases. For the low SBR case, gradients of the low SBR class were backpropagated to obtain the activation maps (\figref~\ref{fig:cam}-top). On the other hand, gradients of the high SBR class were used for the high SBR case (\figref~\ref{fig:cam}-bottom). Generally, the activation maps for high SBR prediction highlighted the nigrosome, parts of the substantia nigra, and parts of the cerebral peduncle. Contrarily, the activation maps for low SBR mainly highlighted the red nucleus region with a small minor focus on parts of the substantia nigra. The low SBR activation maps usually did not show any attention near the nigrosome because the nigrosome hyperintensity is lost in the low SBR cases. This coincides with previous medical research \cite{lehericy2017role, bae2016loss} where nigrosome hyperintensity was used as an identifier for normal/abnormal SPECT findings.

The attention regions looked similar in both the standard and proposed models. However, careful investigation of the maps revealed that the proposed regressor gave stronger and better coverage of the target regions. For example, the low SBR heatmaps of the standard regressor did not or partially include the red nucleus, while the proposed regressor better encompassed the red nucleus region. As a result, the prediction of the proposed method is closer to the true values. We also had similar findings for the high SBR maps where the proposed method gave stronger attention to the nigrosome and substantia nigra. Moreover, the symmetry of the input pairs is well-resembled in the activation maps of the proposed method.

\begin{table*}[!t]

\caption{Prediction Uncertainty Comparison.}
\label{tab:interval}
\renewcommand{\arraystretch}{1.2}
\centering
\begin{tabular}{l | c c c | c c c}
\hline
 & \multicolumn{3}{c|}{\bf Optimal (Sharpness,CP)} &\multicolumn{3}{c}{\bf Sharpness @ $95\%$ CP}\\
{\bf Methods} & Right & Left & Avg. & Right & Left & Avg.\\
\hline
MC+VGG & $(\bs{0.815}, 0.809)$ & $(0.802,0.822)$ & $(0.809,0.816)$ & $0.693$ & $0.705$ & $0.699$\\
MC+Proposed & $(0.811, 0.845)$ & $(0.815,0.\bs{840})$ & $(\bs{0.813},0.843)$ & $0.710$ & $0.728$ & $0.719$\\
Symmetric MC+Proposed & $(0.806, \bs{0.867})$ & $(\bs{0.819},0.831)$ & $(\bs{0.813},\bs{0.849})$ & $\bs{0.733}$ & $\bs{0.765}$ & $\bs{0.749}$\\
\hline
\end{tabular}
\end{table*}

\subsection{Uncertainty Analysis}
\label{sec:uncertainty}
Based on the sharpness-CP curve in Fig~\ref{fig:auc}, we comparatively analyze the uncertainty of the regressor models in Table~\ref{tab:interval}. We present our discussion on this in two steps. First, we compare the uncertainty between the standard regressor and the proposed symmetric regressor. Then, we present the advantage of the symmetric MC uncertainty estimation method over the typical MC method. We base our comparison on the sharpness (i.e. precision) and coverage (i.e. recall) measures as mentioned earlier in Section~\ref{sec:res_uncertainty}. Besides presenting the optimal sharpness and coverage, we also compare the sharpness at a high coverage (see Table~\ref{tab:interval}). Specifically, we selected the point of $95\%$ coverage, which is the most commonly used confidence rate.


From Table~\ref{tab:interval}, we can see that the prediction intervals of the proposed regressor had higher optimal sharpness and coverage compared with the standard regressor. The average normalized width of the standard regressor's prediction interval was $0.191$ with a $81.6\%$ probability of covering the true SBR within it. On the other hand, the  symmetric regressor's interval width was $0.187$ with a $84.3\%$ coverage probability. The higher certainty and precision of the proposed regressor was also prominent at the $95\%$ coverage.

The interval estimated by the symmetric MC dropout method had a an optimal width of $0.187$ with a coverage of $84.9\%$, which is only slightly improved from the typical MC method. However, the improved and sharper estimation of the interval is prominent at higher coverage. For example, at $95\%$ coverage, the symmetric MC method gave an average interval of width $0.251$, which is a notable improvement from the standard MC interval width of $0.281$. Therefore, the symmetric regressor coupled with the symmetric MC uncertainty can be used in the clinical sites more reliably.

\section{Conclusion}
\label{sec:conclusion}
We proposed a symmetric regressor for predicting SBR in DAT SPECT based on the nigral MRI patch. Unlike the typical regressors, the symmetric regressor uses a paired input-output model, which takes the right and left nigral patches simultaneously and predicts both the right and left SBRs. Moreover, it uses an additional symmetric loss which enforces that right and left SBRs should be close to each other. This facilitated an improved model training by utilizing the right-to-left correlation (both in the input and output space) given by the symmetric property of the right and left nigrae. Furthermore, we proposed a symmetric MC dropout utilizing the symmetry for predicting the interval of uncertainty in SBR prediction. Experimental results revealed significant improvement of the symmetric regressor over the standard regressors in terms of prediction error and correlation w.r.t. the true SBRs. The learned features of the symmetric regressor were also more powerful and discriminative. Moreover, the regressor had better activation maps with improved attention in the target regions. Finally, the resultant prediction interval given by the symmetric MC dropout also had high sharpness at $95\%$ coverage probability. Therefore, the proposed method can potentially be helpful for safer MRI-based PD monitoring.

\appendix

\section*{Acknowledgement}
Funding: This work was supported in part by Hankuk University of Foreign Studies (HUFS) Basic Research Fund, South Korea (No. 20231140001) and the National Research Foundation of Korea (NRF) grant funded by the Korea government (MSIT) (No. 2022R1F1A1064530) and grant No. 02–2023-0011 from the Seoul National University Bundang Hospital Research Fund (corresponding authors: Il Dong Yun and Yun Jung Bae).

%


\bibliography{mybib}

\end{document}